\documentstyle[12pt]{article}
\begin{document}
\begin{center}
{\Large Asymptotic Approximation for Bethe-Salpeter Equation\\
and its Applications to Deuteron }
\end{center}
\begin{center}
{Susumu Kinpara}
\end{center}
\begin{center}
{\it National Institute of Radiological Sciences, Chiba 263-8555, Japan}
\end{center}
\begin{abstract}
Bethe-Salpeter equation is solved for bound state composed of two fermions mediated by pion exchange force of the pseudovector coupling.
Expanding the amplitude by gamma matrices the one-dimensional integral equation is derived. It reproduces the binding energy of deuteron.
The relation with the quadrupole moment is also discussed in the framework of the asymptotic approximation.
\end{abstract}
\hspace*{4.mm}
Since discovery of pion the meson exchange picture of the nuclear force has been playing an important role in the nuclear many-body problem accompanied with an abundance of heavier mesons.
The nuclear structure calculation with the meson exchange interactions allow us to understand significance of the relativistic framework for nuclear matter and finite nuclei systematically$\cite{Bouyssy}\cite{Kinpara}$.
\\\hspace*{4.mm}
Two-body system composed of proton and neutron that is deuteron nucleus gives opportunity to investigate whether the parameters of the nuclear force are concerned with the nuclear bound state properties or not, without beeing subject to many-body effects such as the exclusion principle by occupied fermions, the medium effects by the one-body potential and so on. 
Particulerly, it is the main purpose to study the observables from the field theoretical point of view and to dicuss implications in the results of the deuteron structure calculations.
\\\hspace*{4.mm}
The procedure to deal with the Bethe-Salpeter equation for the bound state consists of several steps to reach the solution.
The Wick rotation is usually practiced before the equation is expanded by the four-dimensional spherical harmonics.
In the present study we follow a different way to calculate the observables by expanding the Bethe-Salpeter amplitude directly in the Minkowski space-time and performing the Fourier transform to convert it into that of the coordinate representation.
\\\hspace*{4.mm}
The two-particle Green function for fermion satisfies the ladder approximated self-consistent equation
with the non-interacting correspondent one in free space. Assuming the bilinear form and the existence of pole terms in the exact Green function at the bound state, the homogeneous Bethe-Salpeter equation$\cite{Gell-Mann}$ determining the non-perturbative properties of the composite system is derived as follows\\
\begin{equation}
[\gamma\cdot(\frac{P}{2}+p)-M][\gamma\cdot(\frac{P}{2}-p)-M]\chi_P(p)=\int \frac{d^4 p^\prime}{(2 \pi)^4}
V(p-p^\prime)\chi_P(p^\prime)
\end{equation}
in momentum space.
Here $\it P$, $\it p$ and $\it M$ are the total four-momentum, the relative four-momentum and the nucleon mass respectively.
The ladder approximation for the kernel of the interaction is relevant to the subject in the present study
and corrections for the lowest order two-body interaction kernel makes the problem complicated exceedingly.
\\\hspace*{4.mm}
The Bethe-Salpeter amplitude $\chi_P(p)$ is expanded by a set of sixteen 4$\times$4 matrices 
$\Gamma_i(i=1,\cdot\cdot\cdot,16)$ and it is dissolved into scalar, vector, tensor, axial-vector and pseudo-scalar sectors on the
expansion coefficients$\cite{Nakanishi}$.
When the center of mass energy is set equal to zero ($\it P_{\rm 0}\equiv\;$0), equations are split into three parts and one of them is really the single equation$\cite{Goldstein}$.
The zero energy solutions are important to investigate the short-range behavior of the radial wave functions under the 
singular potential.
In general, the property of decoupling does not maintain any more 
with inclusion of the energy term ($\it P_{\rm 0}\ne\;$0) precisely.
\\\hspace*{4.mm}
Illustrating the procedure it is sufficient to take only the zeroth vector component $\chi_0(p)$ and to neglect other amplitudes in the simultaneous equation as
\begin{equation}
(M^2-\frac{P^2}{4}-p^2)\chi_0(p)={\vec\tau_p}\cdot{\vec\tau_n}\,f_\pi^2 \, i \int \frac{d^4 p^\prime}{(2 \pi)^4} 
\frac{\chi_0(p^\prime)}{m_\pi^2-(p-p^\prime)^2-i \epsilon}
\end{equation}
\begin{equation}
P \cdot p \; \chi_0(p)=0
\end{equation}
accompanying the auxiliary relation eq. (3).
It is seen that the existence of eq. (3) restricts regions of the solution within the plane $t$=0
as to the relative time $t$ in the four dimensional configuration space.
\\\hspace*{4.mm}
Since the present choice of the Bethe-Salpeter amplitude has nothing to do with the plane wave solution at the low energy limit, the coupling constant has a different correspondence 
$f_{\pi}^{\rm 2} \leftrightarrow -g_{\pi}^{\rm 2}$ between the pseudovector ($f_{\pi}$) and pseudoscalar ($g_{\pi}$) couplings in the present study.
The minus sign is due to the manipulation of the delta force subtraction in the pseudovector coupling.
We have dropped out the delta-type force by directly differentiating the Feynman propagator.
\\\hspace*{4.mm}
There exists another important feature in the interaction of the pseudovector coupling; the appearance 
of the $\sim$1/$r^4$ force.
The $\sim$1/$r^4$ force is strong at the short-range core region and expected to be effective for calculating the core radius of deuteron so as to reproduce the binding energy.
By taking account of the $\sim$1/$r^4$ force the core radius is fallen into about the half for the tensor sector showing that the $\sim$1/$r^4$ force is the origin of the short-range repulsive core of the nuclear force.
\\\hspace*{4.mm}
It is accustomed to using the cutoff $\Lambda$ to avoid the divergence at the vertex of the meson exchange interactions
in the context of the meson theoretical calculation.
We generalize the cutoff procedure from the 3-vector momentum to the four dimensional correspondent one as
$\Lambda^2/(\Lambda^2-p^2)$
, multiplied to the propagator of pion in the momentum space.
After the Fourier transform it is found that the potential is further tractable since the leading order 
is no longer $\sim$1/$r^4$, changed to $\sim$1/$r^2$ potential appreciably.
\\\hspace*{4.mm}
The Feynman propagator in the configuration space is given by using
the modified Bessel function of the second kind$\cite{Zhang}$ (${\it K}_1(m_\pi\sqrt{-x^\mu x_\mu})$)
expanded at $x^\mu x_\mu=0$ in the present calculation. 
It is remarkable that the leading order of the interaction expressed as the inverse square shape is same as 
that of the exact propagator for massless scalar boson.
Therefore, effects of the higher-order terms are attributed to the mesons endowed with masses by the spontaneous
chiral symmetry breaking$\cite{Walecka}$.
\\\hspace*{4.mm}
By Fourier-transforming equations (1) and (2) and
operating the limit as $t\rightarrow 0$ a differential equation is given accordingly.
The form is unexpectedly equivalent to the eigenvalue problem for the nonrelativistic quantum system under the attractive force.
For the centrifugal force prevents deuteron from acquiring the proper binding energy, the orbital anguler momentum $\it l$ is set to $l=0$ supposing the spatial distribution of the amplitude to be spherically symmetric.
The restriction on the solution is supported by the experimental finding that the electric quadrupole moment of deuteron is too small to calculate it conveniently by introducing the spherical harmonics of the higher-order $l > \rm 0$ unless there is a mixing term to include the D-state among the solution. 
Whereas the vector sector has no room for entering the D-state, the tensor sector possesses the mixing term independent of the meson exchange interaction as we will see it afterwards.
\\\hspace*{4.mm}
When only the lowest-order inverse square potential remains the solution of the differential equation is represented in terms of a linear combination of bases $u_\nu^{(j)}(-i \kappa x)$ ($\it j$=1,2) irrespective of the boundary conditions.
Because of the scaling property between the root of the eigenvalue ($\kappa\equiv\sqrt{\scriptsize {\it M |E|}}$) 
and the coordinate ($\it x$) the energy spectrum is known to be continuous.
The higher-order pion mass corrections of the nuclear force signifies the descrete energy spectrum in conclusion.
Here, order of the bases are given by $\nu^{\rm 2}=\frac{\rm 1}{\rm 4}-g$, where the strength of the inverse square potential ${\it g}$ has the upper bound ${\it g}\leq\frac{\rm 1}{\rm 4}$ so as to obtain the binding energy by solving the equation subsequently.
\\\hspace*{4.mm}
The residual higher-order interaction potential $v(\it x)$ improves coefficients of the linear combination and the solution $\psi(x)\equiv u(x)/x$ is given by the self-consistent integral equation$\cite{Nauenberg}$ as\\
\begin{eqnarray}
u(x)=u_\nu^{(1)}(-i \kappa x)[\;A + \frac{1}{2 \kappa} \int_0^x u_\nu^{(2)}(-i \kappa x^\prime)v(x^\prime)u(x^\prime)dx^\prime\;]\nonumber\\
+u_\nu^{(2)}(-i \kappa x)[\;B - \frac{1}{2 \kappa} \int_0^x u_\nu^{(1)}(-i \kappa x^\prime)v(x^\prime)u(x^\prime)dx^\prime\;],
\end{eqnarray}
where the coordinate $\it x$ is in units of inverse of the pion mass $m_\pi$.
The strength of the interaction $g=-\frac{\langle{\vec\tau_p}\cdot{\vec\tau_n}\rangle f_{\pi}^{\rm 2}}{(2\pi)^2}$ is represented with the constant $f_{\pi}$ of the pseudovector coupling and the factor $\langle{\vec\tau_p}\cdot{\vec\tau_n}\rangle=-3$ in front of it is owing to the isospin state $I=0$ assigned to deuteron. 
It is noted that the coupling constant is modified by $\Lambda$
and the other mesons in the actual calculation.
\\\hspace*{4.mm}
For the inverse square potential ($v(\it x)$=0) the constant $\it A$ is equal to zero 
due to the divergent behavior of $u_\nu^{(1)}(-i \kappa x)$ as $\it x\rightarrow\infty$, thus, the discrete spectrum of the binding energy is given by zeros 
of $u_\nu^{(2)}(-i \kappa x)$ at the point to characterize the theory$\cite{Gupta}$.
When the residual interaction is added ($v(\it x)$$\neq$0) the boundary condition to give the zero point of $u(\it x)$
simply determines the ratio $A/B$ instead of $\it \kappa$ at the edge of the core potential.
\\\hspace*{4.mm}
From the nucleon-nucleon scattering experiment it is well known that the two-body nuclear system essentially
needs a core potential.
The boundary condition for the wave function is justified at a short-range region ($x_{\rm 0}$)
whether one chooses the hard-core ($x_{\rm 0}\sim{\rm 0.4 fm}$) or the soft-core ($x_{\rm 0}\sim{\rm 0}$) in the treatment
of the singular potential consequently.
It is noted that the condition is not necessarily indispensable to derive the binding energy in the present scheme 
because the discrepancy in the potential from the simple $\sim$1/$r^2$ shape imposes another condition on the equation at the asymptotic region. 
\\\hspace*{4.mm}
When $v(\it x)$ is included the coefficient of the component $u_\nu^{(1)}(-i \kappa x)$ in eq. (4)
has to be dropped out at $x\rightarrow\infty$ as follows\\
\begin{eqnarray}
A + \frac{1}{2 \kappa} \int_0^\infty u_\nu^{(2)}(-i \kappa x^\prime)v(x^\prime)u(x^\prime)dx^\prime=0.
\end{eqnarray}
The equation on $\it \kappa $ is no longer trivial and becomes feasible for determining the binding energy of deuteron.
In order to solve it value of the constant $B$ is determined up to the multiplicative constant of the Bethe-Salpeter amplitude by using the asymptotic form of $u_\nu^{(2)}(-i \kappa x)$ in eq. (4). 
\\\hspace*{4.mm}
It has been shown that eq. (5) allows only one ground state energy 
appreciably, excluding an incredibly shallow excited state which seems to be difficult to observe exceedingly.
The spectrum of the energy is largely different from that of the inverse square potential.
This implies the accumulation of the energy level is removed 
by the repulsion of the residual interaction $v(\it x)$
and only the ground state remains therein to construct deuteron.
\\\hspace*{4.mm}
The zero energy solutions of the Bethe-Salpeter equation extensively investigated are efficient to analyze 
the spectrum of the energy when properties of the quantum system depend much
on the short-range or the high energy parts of the nuclear force.
In turn, the long-range parts of the nuclear force are paid attention to here.
Therefore, the constant ($\sim O(p^{\rm 0})$) terms such as the total energy of the center of mass system 
are essential to investigate the properties by the asymptotic approximation. 
\\\hspace*{4.mm}
In spite of the tractability the vector sector is yet unsatisfactory
for the gamma matrix correspondent to it does not give the eigenvalue of spin S=1.
It is likely that the above mentioned procedure to derive the binding energy is effective also in the other sectors of the gamma matrices.
On the other hand, the tensor sectors are suitable for the structure of deuteron because the total angular momentum
and the spin is realized by the triad in this sector.
The equation makes possible to add the higher components of the orbital angular momentum, mainly the D-state, expected to reproduce observables in the three-dimensional configuration space.
There exists another reason that the tensor sector is expected to construct the wave function of deuteron.
It is another possible ingredient having
the property of symmetry on the gamma matrices (${}^t$ $\Gamma_T$=$\Gamma_T$) as well as the vector sector, necessary for two identical fermions.
\\\hspace*{4.mm}
Whereas the vector sector does not give the observable on the spin of deuteron, the spin 1 system 
of the eigenfunctions could be built by the tensor sector appreciably.
We select a set of three polar components $T_{0\,i}$ ($\it i$=1,2,3) in the tensor sectors 
to make them act as the 3-vector in the cartesian coordinates.
Expanding it by the vector spherical harmonics $\vec{\rm Y}^J_{M\,(1\,l\,)}(\theta,\phi)$ on the
angular dependence the ${}^3S_1$-${}^3D_1$ system of deuteron is reduced to the simultaneous differential equation
on the radial part of the wave functions $\psi_S(x)$ and $\psi_D(x)$ accordingly.
Unlike the vector sector previously investigated, it is one of the characteristics of the tensor sector 
that there exists the mixing term to connect ${}^3S_1$ and ${}^3D_1$ states in the equation independent of 
the meson-exchange interactions consistently.
Then we proceed our calculation for the electric quadrupole moment of deuteron 
by the asymptotic form of the wave functions, where terms of the interaction are neglected since they decays
faster than the mixing term at the asymptotic region.
\\\hspace*{4.mm}
As well as the vector sector the differential equation for the radial wave function of the tensor sector is also obtained 
and it is solved in the same way.
The core radius 0.3-0.4 fm is achieved such that the binding energy of deuteron is reproduced exactly by adjusting the parameters of the interactions.
The above mentioned four-dimensional cut-off mass $\Lambda$ is chosen a rather small value 
as $\Lambda\sim\;$0.4-0.5 GeV about half as large as that of the traditional meson-exchange interaction.
Nevertheless, it is convinced that the decrease of the repulsive force by changing $\Lambda$ artificially 
to the lower value is compensated with the core potential.
Thus, by making the range approach zero as the binding energy is kept intact, $\Lambda$ could be implemented to go to the appropriate one around 1GeV.
\\\hspace*{4.mm}
The attractive $\sigma$-meson exchange force with the appropriate value of the coupling constant is necessary   
to hold the deuteron under the bound state substantially.
It is seen peculiar to the tensor sector that the vector mesons such as $\omega$ and $\rho$ mesons do not contribute
to the interactions of the equation and the short-range repulsive forces are substituted to the pion-interaction, particularly, the derivative terms with the modification by $\Lambda $ introduced to treat the inverse fourth power potential.
This situation is different from the nuclear many-body systems in which the $\omega$ and $\rho$ mesons serve
as the short-range repulsive force.
\\\hspace*{4.mm}
In the actual calculation of the binding energy we have neglected the mixing term approximately along with the non-diagonal elements of the orbital angular momentum in the interaction of pion. 
Besides the binding energy the quadrupole moment of deuteron is also expected to make the D-state appear
if the above mentioned mixing terms are taken into account exactly. 
Then, we investigate whether the approximation
is consistent with the derivation of the quadrupole moment of deuteron.
\\\hspace*{4.mm}
The radial wave function for the D-state is expressed in the integral form by using the Green function 
for the time-independent static Klein-Gordon operator under the central potential with use of the S-state wave function as the source term.
The higher-order effects originating from the central force potential term is not calculated here on the assumption that the plane wave approximation works well in the asymptotic framework of the wave functions. 
In order to perform the integration, $\psi_S(x)$ and the derivative 
on $\it x$ as $\psi_S(x)\sim{\rm exp}[-\alpha \kappa x]/x$ and $\psi^\prime_S(x)\sim-\alpha \kappa\,{\rm exp}[-\alpha \kappa x]/x$ 
are required at the asymptotic region $x\rightarrow\infty$ with the coefficient $\alpha$ determined self-consistently.
\\\hspace*{4.mm}
The D-state probability of deuteron is extracted from the experimentally measured value of the magnetic dipole moment 
as $P_D$=0.039 and the normalization condition ($P_S$+$P_D$=1) is applied simply. 
It is verified on the basis of the standard value $P_D$ that the magnitude of the mixing term is too large to give the most desirable form of the asymptotic wave function ($\alpha$=1).
Therefore, effects of the other components are required to correct the existing mixing term.
\\\hspace*{4.mm}
The additional mixing terms are supplied by the pseudo-scalar and the axial-vector components in the equation
of the tensor sector.
Either of these components alone does not succeed to give the suitable magnitude on the mixing term.
In order to reduce the whole magnitude and obtain the quadrupole moment of deuteron, both of the equations on the components have to be solved simultaniously.
Here, to show the magnitude of the mixing term a parameter $\lambda$ is introduced and  
multiplied to the original mixing term.
It is given by $\lambda = 1$ when there are no effects from the two components.
Thus, including the additional effects $\lambda$ is changed and determined by the following self-consistent equation
as\\
\begin{eqnarray}
\lambda = 1-\frac{3-2 \lambda}{3- \lambda-\frac{4\lambda^2\alpha^4}{3-(3-4\lambda)\alpha^2}},
\end{eqnarray}
the second term on the right hand side gives the joint effects 
of the pseudo-scalar and the axial-vector components.
By the relation between $\psi_S(x)$ and $\psi_D(x)$, $\alpha$ is determined as ${\alpha} = 1$ 
consistently and which concludes $\lambda = 0$ by eq. (6).
This result is anticipated when the binding energy has been derived by neglecting the mixing terms in the analysis of the tensor sector.
\\\hspace*{4.mm}
The mixing term of the tensor sector makes a contribution to determine the value of the quadrupole moment of deuteron whereas details of the meson-exchange interaction is independent of that in the present framework of the asymptotic approximation.
Using the mixing parameter $\lambda$ in the equation of the tensor sector we have found 
that the calculated value of the quadrupole moment converges the optimum value when $\lambda\rightarrow +0$.
It is estimated about 17 percent lower in comparison of the experimental value 0.286$\,{\rm fm^2}$$\cite{Pavanello}$.
This discrepancy is attributed to the asymptotic approximation employed here because the decrease of the wave function around the short-range core region is not taken into account completely.
It is interesting to examine the inner parts of the wave function and to investigate 
the relation between various parameters of the meson-exchange interactions. 

\end{document}